\documentclass[a4paper,reqno,12pt,draft]{article}
\usepackage{amssymb,euscript}
\usepackage{amsmath,amssymb}

\newcommand{\e}{\begin{equation}}
\newcommand{\ee}{\end{equation}}
\newcommand{\ea}{\begin{eqnarray}}
\newcommand{\eea}{\end{eqnarray}}

\newcommand{\bZ}{{\bar Z}}

\begin{document}

\begin{flushright}
\end{flushright}
\begin{flushright}
\end{flushright}
\begin{center}

{\LARGE {\sc Three-Algebras in ${\cal N} = 5, 6$ Superconformal Chern-Simons Theories: Representations and Relations \\ } }

\bigskip
{\sc Jonathan Bagger\footnote{bagger@jhu.edu}}\\
and\\
{\sc George Bruhn\footnote{gbruhn@pha.jhu.edu}} \\
\bigskip

{Department of Physics and Astronomy\\
Johns Hopkins University\\
3400 North Charles St.\\
Baltimore, MD 21218, USA\\}

\end{center}

\bigskip
\begin{center}
{\bf {\sc Abstract}}
\end{center}

In this work we present 3-algebraic constructions and representations for three-dimensional ${\cal N} = 5$ supersymmetric Chern-Simons theories, and show how they relate to theories with additional supersymmetries.   The ${\cal N} = 5$ structure constants give theories with Sp($2N$) $\times$ SO($M$) gauge symmetry, as well as more exotic symmetries known from gauged supergravity.  We find explicit lifts from ${\cal N} = 6$ to 8, and ${\cal N} = 5$ to 6 and 8, for appropriate gauge groups.

\newpage
\section{\label{sec:level1}\sl Introduction}

Over the past few years, there has been an explosion of interest in three-dimensional supersymmetric Chern-Simons gauge theories.  Much progress was sparked by the ${\cal N} = 8$ theory put forth in \cite{Bagger:2006sk},\cite{Bagger:2007jr},\cite{Bagger:2007vi}, and independently in \cite{Gustavsson:2007vu}, that was proposed to describe the world volume theory of coincident M2-branes \cite{Basu:2004ed,Schwarz:2004yj}. \\

The theory contains 8 scalars, $X^I$, which take values in the transverse space, and a 16-component real fermion $\Psi$, which is a two-component real $d=3$ spinor in one of the 8-dimensional spinor representations of the SO(8) R-symmetry group; the supersymmetry parameter $\epsilon$ is in the other.  The fields take values in a 3-algebra, defined by a totally antisymmetric triple product, given by 
\begin{equation}
[T^a ,T^b , T^c] = f^{abc}{}_d T^d.
\end{equation}
The invariant, symmetric inner-product $(T^a ,T^b) = h^{ab}$ raises and lowers indices, so that $f^{abcd}$ is real and totally antisymmetric.  The theory is gauged, with gauge field
\begin{equation}
\tilde{A}_\mu{}^a{}_d = f^{abc}{}_d A_{\mu bc}.
\end{equation}
The gauge field is constrained, so the degrees of freedom balance between bosons and fermions.  The 3-algebra satisfies the so-called fundamental identity,
\begin{eqnarray}
[T^a ,T^b ,[T^c , T^d , T^e ]] &=&  [[T^a , T^b , T^c ],T^d ,T^e]+ [T^c ,[T^a ,T^b ,T^d],T^e] \\[1mm]
\nonumber && +\ [T^c ,T^d ,[T^a ,T^b ,T^e]],
\end{eqnarray} 
which implies that the gauge transformations act as derivations.  These constraints define the ${\cal N} = 8$ theory, of which there is only one (unitary) example: $f^{abcd}\sim\varepsilon^{abcd}$ and $h^{ab}\sim\delta^{ab}$, for which the gauge group is SO(4). \\

More theories can be found by reducing the number of supesymmetries.  These include the ABJM theories, with ${\cal N} = 6$ supersymmetry and U($N$) $\times$ U($N$) gauge symmetry \cite{Aharony:2008ug}, and the ABJ theories \cite{Aharony:2008gk}, with ${\cal N} = 6$ and U($N$) $\times$ U($M$) gauge symmetry, as well as ${\cal N} = 5$ with Sp(2$N$) $\times$ O($M$).  Similar theories were constructed in \cite{Hosomichi:2008jb}.  A classification of the possible ${\cal N} = 6$ theories of ABJM-type was presented in \cite{Schnabl:2008wj}.  \\

None of these constructions made use of a 3-algebra, so it is natural to ask whether they play any role in theories with ${\cal N} < 8$.  In fact, the most general ${\cal N} = 6$ theory was constructed from a 3-algebra in \cite{Bagger:2008se}.  One realization gives rise to an ${\cal N} = 6$ theory with SU($N$) $\times$ SU($N$) gauge symmetry; another describes the ${\cal N} = 6$ U($N$) $\times$ U($M$) ABJ theories.  It has recently been shown that the SU($N$) $\times$ SU($N$) theory is related to the U($N$) $\times$ U($N$) ABJM theory \cite{Lambert:2010ji}, so the 3-algebraic approach indeed describes the complete set of ${\cal N} = 6$ ABJM and ABJ theories. \\

Given these results, one would also like to know the role that 3-algebras play in ${\cal N}=5$ theories.  The quaternionic unitary 3-algebras were classified in \cite{deMedeiros:2008zh}, where it was found that they are in one-one correspondence with the ${\cal N}=5$ Chern-Simons theories presented in \cite{Hosomichi:2008jb} and \cite{Bergshoeff:2008bh}.  In this paper we take a more direct approach and construct the most general three-dimensional ${\cal N} = 5$ superconformal Chern-Simons theories from first principles.  We work in components and close the supersymmetry transformations on the fields.  We find that the theories depend on real structure constants with four upstairs indices, satisfying ${\cal N} = 5$ versions of the fundamental identity.  When the structure constants obey $f^{abcd} = -f^{bacd} = f^{cdab}$, they give rise to ${\cal N} = 5$ truncations of ${\cal N} = 6$ theories, with supersymmetry transformations given in \cite{Bagger:2008se}.  When they obey $g^{abcd} = g^{bacd} = g^{cdab}$, with $g^{(abc)d} = 0$, the theories are purely ${\cal N} = 5$.  For this case, our ${\cal N} = 5$ transformation laws agree with those presented in \cite{Chen:2009cwa}.  Our results are in accord with the classification derived in \cite{deMedeiros:2008zh}.  In addition, they clarify the connection between ${\cal N} = 5$ and ${\cal N} = 6$ theories and show that they both arise as independent solutions to a single set of constraints. \\

In what follows we also present explicit 3-algebra representations for various ${\cal N} = 5$ theories.  We recover all the examples discussed in \cite{Hosomichi:2008jb,Bergshoeff:2008bh,deMedeiros:2008zh}.  We find an Sp(2$N$) $\times$ SO($M$) theory of ABJ-type, with matter fields transforming in the bifundamental representation of the gauge group, as well as an SO(4) $\times$ SU(2) theory with one free parameter.  We also find more exotic theories with gauge groups G$_2$ $\times$ SU(2), with bifundamental matter, and SO(7) $\times$ SU(2), with matter in the 8-dimensional spinor representation of SO(7).  These theories can also found using the ``embedding tensor" approach to $d=3,\ {\cal N} = 8$ gauged supergravity in the conformal limit \cite{Bergshoeff:2008cz,Bergshoeff:2008ix}, or using ${\cal N} = 1$ superspace, as was done in \cite{deMedeiros:2009eq}.\\

Finally, in this paper we also show how to lift certain theories with ${\cal N} = 5$ and ${\cal N} = 6$ supersymmetry to ${\cal N} = 6$ and ${\cal N} = 8$.  We first lift the ${\cal N} = 6$ theory with SU(2) $\times$ SU(2) $\simeq$ SO(4) gauge symmetry to ${\cal N} = 8$.  We then lift the Sp($2N$) $\times$ SO(2) invariant ${\cal N} = 5$ theory to ${\cal N} = 6$.  As a third example, we lift the ${\cal N} = 5$ theory with SO(4) $\times$ SU(2) gauge symmetry to ${\cal N} = 6$ at one point in its parameter space.  At that point, the gauge symmetry is reduced to SO(4) = SU(2) $\times$ SU(2), as required for ${\cal N} = 6$ supersymmetry. \\ 

The layout of the paper is as follows.  In the next section, we review the 3-algebraic construction of the ${\cal N} = 6$ theories.  We present specific representations of the various gauge groups that arise, and we demonstrate the lift to ${\cal N} = 8$.  We then turn our attention to ${\cal N} = 5$ and construct the most general theory based on a 3-algebra.  We find the fundamental identity, and solve it in terms of structure constants of two different kinds.  We discuss explicit representations, and present the lifts from ${\cal N} = 5$ to ${\cal N} = 6$.                           
    
\section{\label{sec:level2}\sl Review of the ${\cal N}= 6$ Construction}
  
In this section, we review the relevant features of the construction in \cite{Bagger:2008se}.  We start by decomposing the SO(8) global symmetry into SO(6) $\times$ SO(2) = SU(4) $\times$ U(1).  The matter fields are a scalar $Z^A _a$ and a spinor $\Psi_{Aa}$, both with U(1) charges $+1$, together with their conjugates $\bar{Z}^a _A$ and $\Psi^{A a} $, where $A=1,...4$ is the SU(4) index and $a$ spans a representation of some gauge group. The 3-algebra structure constants $f^{ab}{}_{cd}$ are no longer necessarily real or totally antisymmetric, but satisfy $f^{ab}{}_{cd} = -f^{ba}{}_{cd} = f^{ba}{}_{dc} = f^{*}_{cd}{}^{ab}$.  The six supersymmetry parameters $\varepsilon^{AB}$ are antisymmetric in $A$ and $B$, and obey the reality condition
\begin{equation}
\varepsilon^{AB} = \frac{1}{2}\varepsilon^{ABCD}\varepsilon_{CD}.
\end{equation}

The ${\cal N}$ = 6 supersymmetry transformations on the scalar and the fermion are
\begin{eqnarray}\label{Neq6trans}
 \nonumber \delta Z^A _d &=& i\bar{\varepsilon}^{AD}\Psi_{Dd}, \\[1mm]
 \delta\Psi_{Dd} &=& \gamma^{\mu}\varepsilon_{AD}D_{\mu}Z^A _d \\[1mm]
\nonumber &&+\  f^{ab}{}_{cd}Z^A _a Z^B _b \bar{Z}_A^c \varepsilon_{BD} + f^{ab}{}_{cd}Z^A _a Z^B_b \bar{Z}_D^c \varepsilon_{AB},
\end{eqnarray}
where the gauge-covariant derivative on the scalar is defined by
\begin{equation}
D_{\mu} Z^A _d = \partial_{\mu}Z^A _d - \tilde{A}_\mu{}^a{}_{d}Z^A _a .
\end{equation}

The transformations on the scalar close according to the supersymmetry algebra,
\begin{equation}
[\delta_1 , \delta_2]Z^A _d = v^{\mu}D_{\mu}Z^A _d + \tilde\Lambda ^a{} _d Z^A _a ,
\end{equation} 
where 
\begin{equation}
v^{\mu} =\frac{i}{2} \bar{\varepsilon}^{CD}_2 \gamma ^{\mu}\varepsilon_{1CD}
\end{equation}
and
\begin{equation}
\label{lambda}
\tilde\Lambda^a {}_d = i\bar{\varepsilon}^{CE}_{[2}\varepsilon_{1]BE}
\,f^{ab}{}_{cd}Z^B_b \bar{Z}^c_C,
\end{equation}
where the antisymmetrization is done without a factor of $\frac{1}{2}$.\\

The transformations on the fermions close similarly,
\begin{equation}
[\delta_1 , \delta_2]\Psi_{Dd} = v^{\mu}D_{\mu}\Psi_{Dd} + \tilde\Lambda ^a{}_d \Psi_{Da},
\end{equation}
provided the equations of motion are satisfied:
\begin{eqnarray}
\label{FermEOM}
 E_{Dd} & = & \gamma ^{\mu} D_{\mu} \Psi_{Dd} -2 f^{ab}{}_{cd}  \Psi_{Ba}Z^B_b \bar{Z}^c _D  \\[1mm]
  \nonumber   & & + \ f^{ab}{}_{cd}\Psi_{Da}Z^B_b \bar{Z}^c_B + \varepsilon_{ABCD}f^{ab}{}_{cd}\Psi^{Cc}Z^A_a Z^B _b \ = \ 0. 
\end{eqnarray}

Finally, the gauge field transformations 
\begin{equation}
\delta \tilde{A}_{\mu}{}^a {}_d = -i f^{ab}{}_{cd} (\bar{\varepsilon}^{BC}\gamma _{\mu} \Psi_{Bb}\bar{Z}^c _C +
\bar{\varepsilon}_{BC}\gamma_{\mu}\Psi^{Cc}Z^{B} _b )
\end{equation}
close as follows,
\begin{equation}
[\delta_1,\delta_2]\tilde{A}_\mu{}^a{}_d =
D_\mu(\tilde\Lambda{}^a{}_d)+
v^\nu \tilde{F}_{\mu\nu}{}^a{}_d + {\cal O}(Z^4) ,
\end{equation}
provided the field strength obeys the following condition:
\begin{eqnarray}\label{Fcondition}
\tilde F_{\mu\nu}{}^a{}_d &=&
 -\partial_\mu \tilde
A_\nu{}^a{}_d+\partial_\nu \tilde A_\mu{}^a{}_d +  \tilde
A_\nu{}^a{}_b\tilde A_\mu{}^b{}_d- \tilde A_\mu{}^a{}_b\tilde
A_\nu{}^b{}_d \nonumber\\[1mm]
&=& -\varepsilon_{\mu\nu\lambda}\left(D^\lambda Z^B_b \bZ^c_B-
Z^B_b D^\lambda \bZ^c_B
-i\bar\Psi^{Bc}\gamma^\lambda\Psi_{Bb}\right)f^{ab}{}_{cd}.
\end{eqnarray}
Canceling the ${\cal O}(Z^4)$-terms leads to the ${\cal N}=6$ fundamental identity,
\begin{equation}
f^{ef}{}_{gb} f^{cb}{}_{ad}+f^{fe}{}_{ab} f^{cb}{}_{gd} +
f^{*}_{ga}{}^{fb} f^{ce}{}_{bd} +f^{*}_{ag}{}^{eb} f^{cf}{}_{bd} =0.
\end{equation}
The fundamental identity ensures that the gauge transformation acts as a derivation.  With these ingredients, it is not hard to construct the ${\cal N}=6$ Lagrangian, written in terms of the 3-algebra.  In the next section, we discuss representations of the ${\cal N} = 6$ gauge groups.

\section{\label{sec:level3}\sl ${\cal N} = 6$ Representations}
 
A representation of the 3-algebra can be constructed from rectangular $M \times N$ matrices, $X,Y,Z,$ as follows:
\begin{equation}
\label{Neq6Mat}
[X,Y;Z] = X Z^{\dagger} Y - Y Z^{\dagger}X,
\end{equation} 
where $Z^{\dagger}$ is the conjugate transpose of $Z$.  This can be interpreted as a gauge transformation on $X_{dl}$, acting via left and right multiplication, with $X$ carrying bifundamental indices $d$ and $l$,
\begin{eqnarray}
\nonumber \delta X_{dl}& =& [X,Y;Z]_{dl} \\[1mm]
   & = &X_{dk}Z^{\dagger kb}Y_{bl} - Y_{dk}Z^{\dagger kb} X_{bl}.
\end{eqnarray}
In this case, the 3-algebra structure constants are given by 
\begin{equation}\label{UcrossU}
f^{aibj}{}_{ckdl} = \delta ^a _d \delta ^b _c\delta ^i _k \delta ^j _l -  \delta ^a _c \delta ^b _d\delta ^i _l \delta ^j _k .
\end{equation}
The structure constants have the correct symmetries and satisfy the ${\cal N}=6$ fundamental identity.\\

Using (\ref{lambda}), it is a simple matter to determine the gauge theories that are constructed in this way.  For this particular 3-algebra, we find
\begin{equation}
\delta Z^A _{dl} = \tilde\Lambda ^{ai} {}_{dl} Z^A _{ai} =
 i\bar\varepsilon^{CE}_{[2}\varepsilon_{1]BE}Z^B_{bl}\bar{Z}_C^{bk}Z^A_{dk} 
- i\bar\varepsilon^{CE}_{[2}\varepsilon_{1]BE}Z^B_{dj}\bar{Z}_C^{cj}Z^A_{cl},
\end{equation} 
The matrix  $\tilde\Lambda ^{ai} {}_{dl}$ is anti-Hermitian, with a nonvanishing trace for $M \neq N$ and a vanishing trace for $M = N$.  As expected, these ${\cal N} = 6$ theories have U($N$) $\times$ U($M$) and SU($N$) $\times$ SU($N$) gauge symmetry.  The original U($N$) $\times$ U($N$) ABJM model can be recovered by gauging the global U(1) symmetry, as was done in \cite{Lambert:2010ji}.  \\

A second choice of structure constants is given by
\begin{equation}
f^{ab}{}_{cd} = J^{ab} J_{cd} + (\delta^a _c \delta^b _d - \delta^a _d \delta^b _c ),
\end{equation} 
where $J^{ab} = i(\sigma^2 \otimes {\bf I}_{N\times N})^{ab}$ is the antisymmetric invariant tensor of Sp(2$N$).  The $f^{ab}{}_{cd}$ also obey the fundamental identity and have the correct symmetries.  As before, we close the algebra to find the gauge transformation on $Z^A _d$,
\begin{eqnarray}
\delta Z^A _{d}\ =\ \tilde\Lambda ^{a} {}_{d} Z^A _{a} &=&  i \bar{\varepsilon}_{[2 } ^{CE} \varepsilon_{ 1]BE}(Z^B _d \bar{Z}^a _C + J^{ab}J_{cd} Z^B _b \bar{Z}_C ^c )Z^A _a \nonumber \\[1mm]
&&  - \ i \bar{\varepsilon}_{[2} ^{CE} \varepsilon_{1]BE}Z^B _b \bar{Z}_C ^bZ^A_d .
\end{eqnarray}  
This transformation is a sum of two parts.  The first is of the form $\delta'Z^A_d = \tilde\Lambda'^a{}_d Z^A_a$; the second is a phase.   It is easy to see that $J_{ab}\Lambda'^b{}_c J^{cd} = \Lambda'^d{}_a $, so the gauge group is simply Sp(2$N$) $\times$ U(1). 

\section{\sl Lift: ${\cal N} = 6 \rightarrow {\cal N} = 8$}\label{6to8}

{}From the above construction, it is possible to find an explicit lift from the $\cal{N}$ = 6 theory with SU(2) $\times$ SU(2) gauge symmetry to the unique ${\cal N} = 8$ theory.  We begin by writing the matter fields $Z^A _{\alpha \dot{\alpha}}$ in SO(4) notation,
\begin{equation}
Z^A _d =  Z^A _{\alpha \dot{\alpha}}\bar{\sigma}_d ^{\dot{\alpha} \alpha},
\end{equation}   
using the ordinary Pauli matrices of \cite{Wess:1992cp} (except taking $\sigma^0 \rightarrow i\sigma^0 = i\bar\sigma^0$ to make the gauge space Euclidean).  Because of the well-known identity
$$(\bar{\sigma}^a \sigma ^b \bar{\sigma}^c - \bar{\sigma}^c \sigma ^b \bar{\sigma}^a)^{\dot{\alpha}\alpha} = -2 \varepsilon^{abcd}\bar{\sigma}_d ^{\dot{\alpha}\alpha},$$
the representation of the SU(2) $\times$ SU(2) transformation given in (\ref{Neq6Mat}) exactly reproduces the 3-algebra of the ${\cal N} = 8$ theory, with $f^{abcd} =\varepsilon^{abcd}$ (we absorb the constant of proportionality into $\varepsilon^{abcd}$). \\

In this notation, we start with the original ${\cal N} = 6 $ supersymmetry transformations presented above, parametrized by $\varepsilon^{AB},$ and construct two additional supersymmetries, parametrized by a {\it complex} spinor $\eta$ of global U(1) charge $+2$.  The most general supersymmetry transformations consistent with these assignments are 
\begin{eqnarray}
\nonumber \delta Z^A _d &=& i\bar{\varepsilon}^{AD}\Psi_{Dd} + i \Theta_1 \bar{\eta}\Psi^A _d \\[2mm]
\delta \Psi_D ^d & = & \gamma ^{\mu}\varepsilon_{AD} D_{\mu}Z^{Ad} + \Theta_2 \gamma ^{\mu}\eta D_{\mu}\bar{Z}^d _D \\[1mm]
\nonumber                && +\ \varepsilon ^{abcd} Z^A_a Z^B _b \bar{Z}_{Dc}   \varepsilon_{AB} - \varepsilon^{abcd}Z^A_a Z^B_b \bar{Z}_{Bc}\varepsilon_{AD} \\[1mm]
 \nonumber    && -\ \Theta_3 \varepsilon^{abcd} Z^A_a \bar{Z}_{Ab}\bar{Z}_{Dc}\eta + \Theta_4 \varepsilon_{ABCD}\varepsilon^{abcd}\eta^* Z^A _a Z^B _b Z^C _c  ,
\end{eqnarray}
for some complex numbers $\Theta_1 ,\Theta_2 ,\Theta_3 ,\Theta_4$.  Note that since the gauge group is SO(4), the gauge indices can be raised and lowered at will.\\
 
Imposing the supersymmetry algebra on the scalar transformation leads to $\Theta_1 = \Theta_3$ and $\Theta_1= \Theta_2$.  In particular, we find 
\begin{equation}
[\delta_1 , \delta_2 ]Z^{A}_d = v^{\mu}D_{\mu}Z^{A}_d + \tilde\Lambda^{a}{}_{d}Z^A _a , \\
\end{equation}
where
\begin{equation}
v^{\mu} = \frac{i}{2} \bar{\varepsilon}^{BC} _2 \gamma^{\mu} \varepsilon_{1BC} + i |\Theta_1|^2 \bar{\eta}_{[2} \gamma^{\mu}\eta^* _{1]}
\end{equation}
and
\begin{eqnarray}
\nonumber \tilde\Lambda^{ad} &=&  i\bar{\varepsilon}^{CE}_{[2}\varepsilon_{1]BE} \varepsilon^{abcd}Z^{B}_b \bar{Z}_{Cc} +3i\Theta_4 \bar{\varepsilon}_{[2 BC}\eta^* _{1]}\varepsilon^{abcd}Z^B _b Z^C _c  \\[1mm]
  && +\ i\Theta_1 \bar{\eta}_{[2}\varepsilon^{BC}_{1]}\varepsilon^{abcd}\bar{Z}_{Bb}\bar{Z}_{Cc} + i|\Theta_1|^2 \bar{\eta}_{[2}\eta^* _{1]}\varepsilon^{abcd} Z^B_b \bar{Z}_{Bc} .
\end{eqnarray}
 Anti-Hermicity of the generator $\tilde\Lambda^{ad}$ requires $\Theta_1 = -3\Theta_4^*$.  This leaves only $\Theta_1$ independent; it can be absorbed into the parameter $\eta$.  \\
 
With these results, the supersymmetry transformations are
 \begin{eqnarray}\label{deltaMAT}
\nonumber \delta Z^A _d &=& i\bar{\varepsilon}^{AD}\Psi_{Dd} + i \bar{\eta}\Psi^A _d \\[2mm]
 \nonumber    \delta \Psi_D ^d & = & \gamma ^{\mu}\varepsilon_{AD} D_{\mu}Z^{Ad} + \gamma ^{\mu}\eta D_{\mu}\bar{Z}^d _D \\[1mm]
\nonumber                && +\ \varepsilon ^{abcd} Z^A_a Z^B _b \bar{Z}_{Dc}   \varepsilon_{AB} - \varepsilon^{abcd}Z^A_a Z^B_b \bar{Z}_{Bc}\varepsilon_{AD} \\[1mm]
&& -\ \varepsilon^{abcd} Z^A_a \bar{Z}_{Ab}\bar{Z}_{Dc}\eta - \frac{1}{3} \varepsilon_{ABCD}\varepsilon^{abcd}\eta^* Z^A_a Z^B _b Z^C _c  .
\end{eqnarray}
Closing on the fermion gives 
\begin{eqnarray}
  \nonumber [\delta_1 , \delta_2]\Psi_{Dd} & = & v^{\mu}D_{\mu}\Psi_{Dd} + \tilde\Lambda^a{}_{d}\Psi_{Da} \\[1mm]
     && +\ \frac{i}{2}\bar{\varepsilon}^{CB} _{[2}\varepsilon_{1]CD}E_{Bd} -\frac{i}{4}\bar{\varepsilon}^{BE} _{2}\gamma ^{\mu}\varepsilon_{1BE}\gamma_{\mu}E_{Dd} \\ 
 \nonumber    && +\ i \bar{\eta}_{[2}\varepsilon_{1]CD}E^C _d - \frac{i}{2}(\bar{\eta} _{[2} \eta^* _{1]} + \bar{\eta}^* _{[2} \gamma^{\mu}\eta_{1]} \gamma_{\mu})E_{Dd}, 
\end{eqnarray}
as required, where $E_{Dd}$ denotes the fermion equation of motion (\ref{FermEOM}).  The same calculation also fixes the transformation of the gauge field,
\begin{eqnarray}\label{deltaA}
\nonumber \delta \tilde{A}_{\mu}{}^{ad} &=& -i\varepsilon^{abcd}\bar\varepsilon_{BC}\gamma_{\mu}\Psi^B_b Z^C_c - i\varepsilon^{abcd}\bar\varepsilon^{BC}\gamma_{\mu}\Psi_{Bb}\bar{Z}_{Cc} \\[1mm]
&& +\ i\varepsilon^{abcd} \bar\eta^* \gamma_{\mu}\Psi_{Bb}Z^B_c  + i\varepsilon^{abcd}\bar\eta\gamma_{\mu}\Psi^B_b \bar{Z}_{Bc}.  
\end{eqnarray}
Closing on $\tilde{A}_{\mu}{}^{ad}$ imposes the constraint (\ref{Fcondition}). \\

The above transformations are manifestly SU(4) $\times$ U(1) covariant.  However, they must also be covariant under SO(8), the ${\cal N} = 8$ R-symmetry group.  As a check, therefore, we compute their transformations under the twelve remaining generators of SO(8)/(SU(4) $\times$ U(1)), which we denote $g^{AB}$, with U(1) charge 2.  The transformations are
\begin{eqnarray}\label{GTT}
\nonumber \delta Z^A_a &=& g^{AB}\bar{Z}_{Ba} \\
\nonumber \delta \Psi_{Ba} &=& -\frac{1}{2}\varepsilon_{BCDE}g^{DE}\Psi^C_a \\
 \delta \varepsilon^{AB} &=& g^{AB}\eta^* + \frac{1}{2}\varepsilon^{ABCD}g^*_{CD}\eta \\
\nonumber\delta \eta &=& -\frac{1}{2}g^{AB}\varepsilon_{AB},
\end{eqnarray}
consistent with the fact that $Z^A_a$, $ \Psi_{Bb}$ and $\varepsilon^{AB}$ live in different SO(8) representations.
The transformations (\ref{GTT}) close into SU(4) $\times$ U(1), as required by the SO(8) algebra.  Moreover, it is not hard to show that the supersymmetry transformations (\ref{deltaMAT}) and (\ref{deltaA}) are covariant under (\ref{GTT}), as they must be.  Thus, for the case of SO(4) gauge symmetry, the supersymmetry transformations (\ref{deltaMAT}) and (\ref{deltaA}) do indeed lift the ${\cal N} = 6$ theory to ${\cal N} = 8$.

\section{\label{sec:level4}\sl ${\cal N} = 5$ Construction}

In this section, we proceed along similar lines to construct the most general ${\cal N}=5$ theories that make use of a 3-algebra.  We start by decomposing the SO(8) global symmetry into SO(5) $\times$ SO(3) $=$ Sp(4) $\times$ SU(2).  We take the eight scalar fields to have the index structure $X^A_{id}$, where $A = 1,..,4$ and $i=1,2$ are indices that refer to the Sp(4) R-symmetry and the global SU(2), respectively; the index $d$ spans a representation of the gauge group.  The Sp(4) indices are raised or lowered with the Sp(4)-invariant tensor, 
$$\omega^{AB} = i(\sigma^2\otimes{\bf I}_{2\times 2})^{AB},$$ 
for which $\omega^{AB}\omega_{BC} = -\delta^A_C$.  Here and elsewhere we adopt the convention $X^A = \omega^{AB} X_B$, $X_A = - \omega_{AB} X^B$ for any symplectic structure.  The supersymmetry parameters are real spinors $\xi^{AB}$, antisymmetric in $A$ and $B$ and traceless, 
\begin{equation}
\omega_{AB}\xi^{AB} = 0,
\end{equation}
so the $\xi^{AB}$ are in the $\bf 5$ of Sp(4).  The superpartner fermions are real spinors as well, with index structure $\Psi_{Aid}$.\\

The most general supersymmetry transformations are of the following form,
\begin{eqnarray}
 \delta X^A_{id} &=& i \bar\xi^{AD} \Psi_{Did} \\
 \nonumber \delta \Psi_{Dld} & =& \gamma^{\mu}\xi_{AD}D_{\mu}X^A_{
ld} \\[1mm]
\nonumber &&+\ \omega_{BD} \xi_{AC} \epsilon^{jk}( f^{abc}{}_d X^A_{l
a} X^B_{jb} X^C_{kc} 
+ g^{abc}{}_d X^A_{ka} X^B_{lb} X^C_{jc}) \\[1mm]
\nonumber &&+\ \omega_{AC} \xi_{BD} \epsilon^{jk}( h^{abc}{}_d X^A_{l
a} X^B_{jb} X^C_{kc} 
+ j^{abc}{}_d X^A_{ka} X^B_{lb} X^C_{jc}) ,
\end{eqnarray}
where the Levi-Civita tensor $\epsilon^{ij}$ raises and lowers the SU(2) indices.   Without loss of generality, we may take $g^{abc}{}_d$ and $j^{abc}{}_d$ to be symmetric in $a$ and $c$.\\

The tensors $g^{abc}{}_d,$ $h^{abc}{}_d$, and $j^{abc}{}_d$ are fixed by closing the supersymmetry algebra on the scalar,
\begin{equation}
[\delta_1,\delta_2]X^A_{id} = v^{\mu}D_{\mu}X^A_{id} + \tilde\Lambda^a{}_d X^A_{ia},
\end{equation}
with $v^{\mu} = \frac{i}{2}\bar\xi^{BC}_2\gamma^{\mu}\xi_{1BC}$.  We find
\begin{equation}
f^{abc}{}_d = 2 g^{cab}{}_d = h^{acb}{}_d = 2j^{cab}{}_d,
\end{equation}
which implies
\begin{equation}\label{vanishinggauge}
\tilde\Lambda^a{}_d X^A_{ia} = \frac{i}{2}\epsilon^{jk}
 \bar\xi^{EF}_{[2}\xi_{1]CF}\omega_{EB}f^{abc}{}_d X^B_{jb}X^C_{kc}X^A_{ia}.
\end{equation}
Because of conflicting symmetries, $\tilde\Lambda^a{}_d$ vanishes, so no gauge transformation appears in the closure of the algebra.  \\

With these conditions, the fermion supersymmetry transformation becomes
\begin{eqnarray}\label{vanishingferm}
\nonumber \delta\Psi_{Dld} &=& \gamma^{\mu} \xi_{AD} D_{\mu}X^A_{ld} \\[1mm]
\nonumber && +\ \epsilon^{jk}(\omega_{BD}\xi_{AC} + \omega_{AC}\xi_{BD}) \\[1mm]
&&\times\ (f^{abc}{}_d X^B_{jb}X^C_{kc}X^A_{la} - \frac{1}{2}f^{abc
}{}_d X^B_{la}X^C_{kc}X^A_{jb}).
\end{eqnarray}
Closing this transformation leads to a trivial theory.  All interaction terms cancel in the equation of motion.  Indeed, upon closer inspection, it is possible to show that the interaction terms in the fermion transformation (\ref{vanishingferm}) also vanish, as indeed they must. \\

To find a nontrivial ${\cal N}=5$ theory, we need to impose a less restrictive global symmetry group.  Therefore, in what follows, we will take the global symmetry group to be the R-symmetry group SO(5) $=$ Sp(4).  Since Sp(4) $\subset$ SU(4), we can carry over many results from ${\cal N} = 6$.  \\

We start by examining the supersymmetry parameters.   We write the ${\cal N} = 6$ parameters $\varepsilon^{AB}$ in terms of the ${\cal N} = 5$ parameters $\xi^{AB}$, together with a {\it real} R-symmetry singlet spinor $\eta$, as follows:
 \begin{eqnarray}\label{epsilonhat}
\nonumber \varepsilon^{AB} &=& \xi^{AB} + i \omega^{AB}\eta \\[1mm]
 \varepsilon_{AB} &=&\xi_{AB} - i \omega_{AB} \eta .
\end{eqnarray} 

In an ${\cal N} = 5$ theory, the Sp(4) indices are raised and lowered using the antisymmetric tensors $\omega^{AB}$ and $\omega_{AB}$, respectively.  For the ${\cal N} = 5$ parameters $\xi_{AB}$, this convention is consistent with the SU(4) R symmetry of the ${\cal N} = 6$ theory:
\begin{eqnarray}
\xi^{AB} &\equiv& \omega^{AC}\omega^{BD} \xi_{CD} \nonumber \\[1mm]
\nonumber  &=&\frac{1}{2}( \omega^{AC}\omega^{BD} -  \omega^{AD}\omega^{BC
} -  \omega^{AB}\omega^{CD}) \xi_{CD} \\[1mm]
 &=&\frac{1}{2}\varepsilon^{ABCD} \xi_{CD}.
\end{eqnarray}
The sign change in the singlet follows from the group theory,
\begin{eqnarray}
\omega^{AB}\eta &\equiv& \omega^{AC}\omega^{BD} \omega_{CD} \eta \nonumber \\[1mm]
\nonumber  &=&-\frac{1}{2}( \omega^{AC}\omega^{BD} -  \omega^{AD}\omega^{BC
} -  \omega^{AB}\omega^{CD}) \omega_{CD} \eta \\[1mm]
 &=&-\frac{1}{2}\varepsilon^{ABCD} \omega_{CD} \eta .
\end{eqnarray}
It is also necessary for the closure of the supersymmetry transformations, as can be checked for the free case.\\
 
We next consider the fields.  The {\bf 4} of Sp(4) is obtained from the {\bf 4} and $\bar {\bf 4}$ of SU(4) by imposing a reality condition.  For the case at hand, we impose the following constraints on the fields of the ${\cal N} =6$ theory:\footnote{Our constraints differ by a critical sign from those in ref.~\cite{Chen:2009cwa}.}
\begin{eqnarray}\label{constraint}
\nonumber \bZ^a_A &=& -J^{ab}\omega_{AB}Z^B_b \\[1mm]
\Psi^{Aa} &=& -J^{ab}\omega^{AB}\Psi_{Bb}.
\end{eqnarray}
Here $\omega_{AB}$ is the antisymmetric Sp(4) invariant tensor, while $J_{ab}$ is an invariant (antisymmetric) tensor of the gauge group, with $J_{ab} J^{bc} = -\delta_a^c$.  The minus sign in the second term renders the constraint consistent with the ${\cal N} = 5$ supersymmetry transformations.  The constraint is inconsistent with the transformation parametrized by $\eta$, so it explicitly breaks ${\cal N} = 6$ supersymmetry to ${\cal N} = 5$. \\

With this constraint, we can write the ${\cal N} = 5$ supersymmetry transformations entirely in terms of the fields $Z^A_a$ and $\Psi_{Dd}$.  The most general transformations take the following form,
\begin{eqnarray}\label{transf1}
\nonumber \delta Z^A_d &=& i\bar{\xi}^{AD}\Psi_{Dd} \\[1mm]
\nonumber \delta\Psi_{Dd} &=& \gamma^{\mu}\xi_{AD}D_{\mu}Z^A_d + f
_1^{abc}{}_dZ^A_aZ^B_bZ^C_c \xi_{DC}\omega_{AB} \\[1mm]
 &&+\ f_2^{abc}{}_d Z^A_aZ^B_bZ^C_c\xi_{AB}\omega_{DC},
\end{eqnarray} 
where, without loss of generality, we take $f_1^{abc}{}_d$ and $f_2^{abc}{}_d$ to be antisymmetric in their first two indices.  Closing on the scalar gives 
\begin{equation}
[\delta_1,\delta_2]Z^A_d = v^{\mu}D_{\mu}Z^A_d + \tilde{\Lambda}^a{}_d Z^A_a,
\end{equation}
with 
\begin{equation}\label{Neq5gauge}
\tilde{\Lambda}^a{}_d = i f_2^{abc}{}_d Z^B_bZ^C_c \omega_{DC} 
\bar{\xi}^{DF}_{[2}\xi_{1]BF}  ,
\end{equation}
where
\begin{equation}
f_1^{abc}{}_d = \frac{1}{2} (f_2^{bca}{}_d - f_2^{acb}{}_d).
\end{equation}
This implies
\begin{eqnarray}\label{transf2}
\nonumber \delta\Psi_{Dd} &=& \gamma^{\mu}\xi_{AD}D_{\mu}Z^A_d -
f_2^{acb}{}_d Z^A_aZ^B_bZ^C_c \xi_{DC}\omega_{AB} \\[1mm]
&&+\ f_2^{abc}{}_d Z^A_aZ^B_bZ^C_c\xi_{AB}\omega_{DC}.
\end{eqnarray}
Closing on the fermion gives
\begin{eqnarray}
\nonumber [\delta_1,\delta_2]\Psi_{Dd} &=& v^\mu D_\mu \Psi_{Dd} +
\tilde\Lambda^a{}_d\Psi_{Da}\\[1mm]
\nonumber &&-\ \frac{i}{2}\bar\xi_{[1}^{AC}\xi_{2]AD}E_{Cd} +\ \frac{i}{4}(\bar\xi^{AB}_1\gamma_\nu\xi_{2AB})\gamma^\nu
E_{Dd}, 
\end{eqnarray}
with the following fermion equation of motion:
\begin{eqnarray}
\nonumber  E_{Dd} &=& \gamma^{\mu}D_{\mu}\Psi_{Dd} \\
\nonumber&& -\ f_2^{abc}{}_d(\Psi_{Dc}Z^A_aZ^B_b + \Psi_{Db}Z^A_aZ^B_c )\omega_{AB} \\
&& +\ 2 f_2^{abc}{}_d (\Psi_{Ab}Z^A_aZ^C_c +
\Psi_{Ac}Z^A_aZ^C_b) \omega_{DC}\ =\ 0.
\end{eqnarray}

With these assignments, the gauge field transforms as
\begin{equation} \label{transf3}
\delta\tilde{A}_{\mu}{}^a{}_d \ =\  -i (f_2^{acb}{}_d + f_2^{abc}{}_d)
\omega^{BE}\bar{\xi}_{EC}\gamma_{\mu}\Psi_{Bb}Z^C_c .
\end{equation}
Closing on the gauge field imposes additional constraints:
\begin{eqnarray}
f_2^{abc}{}_{g}(f_2^{edg}{}_{f} + f_2^{egd}{}_{f})  Z^A_a Z^B_b Z^C_c Z^D_d \omega_{AD}\omega_{BC}
&=& 0 \nonumber \\
f_2^{abc}{}_{g}(f_2^{edg}{}_{f} + f_2^{egd}{}_{f})  Z^A_a Z^B_b Z^C_c Z^D_d \bar\xi_{AB[1} \gamma^\mu\xi_{2]CD} &=& 0.
\label{FIprecursor}
\end{eqnarray}
These two constraints must be satisfied by the ${\cal N}=5$ fundamental identity.\\

Up to now, we have worked in complete generality.  To proceed further, we impose symmetries on the structure constants $f_2^{abc}{}_d$.  The most obvious choice is 
\begin{equation}
f_2^{abcd} = f^{abcd} = - f^{bacd} = f^{cdab} ,
\end{equation}
as in ${\cal N} = 6$.  With this choice, the calculations work out just as before.  In particular, the conditions (\ref{FIprecursor}) 
are satisfied by the ${\cal N}=5$ restriction of the ${\cal N}=6$ fundamental identity:
\begin{equation}\label{primordial}
J_{gj}(f^{abfg}f^{jhcd} + f^{agfd}f^{hbjc} + f^{ahfg}f^{jbdc} + f^{agfc}f^{bhjd}) = 0.
\end{equation}
In this case, the supersymmetry transformations are those of ref.~\cite{Bagger:2008se}.\\

A second and more interesting choice is to take 
\begin{equation}
f_2^{abcd} = g^{acbd} - g^{bcad},
\end{equation}  
where
\begin{equation}
\label{symm1}
g^{acbd} = g^{cabd} = g^{bdac}
\end{equation}  
so $f_2^{abcd}$ has all the right symmetries.  As we shall see, this choice generates ${\cal N} = 5$ theories that are not restrictions of ${\cal N} = 6$.  The conditions (\ref{FIprecursor}) are satisfied if\footnote{We thank Jos\'e Figueroa-O'Farrill and Paul de Medeiros for emphasizing the importance of (\ref{symm2}).}
\begin{equation}
\label{symm2}
g^{(acb)d} = 0
\end{equation}  
and
\begin{equation}\label{Neq5FI}
J_{gj}(g^{afbg}g^{jchd} + g^{afgd}g^{hjbc} + g^{afhg}g^{jdbc} + g^{afgc}g^{bjhd}) = 0.
\end{equation}
This is the ${\cal N} = 5$ fundamental identity, which was also found in \cite{Bergshoeff:2008bh} by taking the conformal limit of three-dimensional gauged supergravity. \\

Substituting $g^{abc}{}_d$ for $f_2^{abc}{}_d$ in (\ref{transf1}), (\ref{transf2}) and (\ref{transf3}), we find the ${\cal N} = 5$ supersymmetry transformations
\begin{eqnarray}
\nonumber \delta Z^A_d &=& i\bar{\xi}^{AD}\Psi_{Dd} \\[1mm]
\nonumber \delta\Psi_{Dd} &=& \gamma^{\mu}\xi_{AD}D_{\mu}Z^A_d - g^{abc}{}_d Z^A_aZ^B_bZ^C_c \xi_{DB}\omega_{AC} \\[1mm]
 &&+\ 2g^{abc}{}_d Z^A_aZ^B_bZ^C_c\xi_{AC}\omega_{DB} \nonumber\\[1mm]
\delta\tilde{A}_{\mu}{}^a{}_d &=&  3i g^{bca}{}_d
\omega^{BE}\bar{\xi}_{EC}\gamma_{\mu}\Psi_{Bb}Z^C_c .
\end{eqnarray} 
These transformations close into a translation and a gauge variation, with parameter
\begin{equation}
\tilde{\Lambda}^a{}_d = - \frac{3i}{2} g^{bca}{}_d Z^B_bZ^C_c \omega_{DC} 
\bar{\xi}^{DF}_{[2}\xi_{1]BF} .
\end{equation}
These are the same transformations that were found, starting from different assumptions, in ref.~\cite{Chen:2009cwa}.

\section{\label{sec:level5}\sl ${\cal N} = 5$ Representations}

In this section we construct ${\cal N} = 5$ gauge theories, built from symmetric structure constants $g^{abcd}$, with gauge transformations
\begin{equation}
\delta Z^A_d = \tilde{\Lambda}^a{}_d Z^A_a =-\frac{3i}{2} g^{bca}{}_d Z^B_bZ^C_c \omega_{DC} 
\bar{\xi}^{DF}_{[2}\xi_{1]BF}  Z^A_a.
\end{equation}
We will see that there are a host of such theories, including some with free parameters or exceptional gauge groups, in stark contrast to ${\cal N} = 6$ or 8.\\

We start by constructing a set of $g^{abcd}$ that lead to an Sp(2$N$) $\times$ SO($M$) gauge group.  There are four combinations of the invariant tensors of Sp(2$N$) and SO($M$) that have the symmetries (\ref{symm1}):
\begin{eqnarray}
g_1^{aibjckdl} &=& (\delta^{ac}\delta^{bd} - \delta^{ad}\delta^{bc})J^{ij}J^{kl} \\
\nonumber g_2^{aibjckdl} &=& (J^{ik}J^{jl} + J^{jk}J^{il})\delta^{ab}\delta^{cd} \\
\nonumber g_3^{(\pm) aibjckdl} &=&(\delta^{ac}\delta^{bd} \pm \delta^{ad}\delta^{bc})(J^{ik}J^{jl} \pm J^{jk}J^{il}),
\end{eqnarray}
where $i,j,... = 1,...\ 2N$ are Sp($2N$) indices, and $a,b,... = 1,...\ M$ are SO($M$).  From them, we must select linear combinations that satisfy (\ref{symm2}) and the fundamental identity (\ref{Neq5FI}).  \\

In fact, there are just two linear combinations that do the job: 
\begin{eqnarray}\label{N5struct}
g^{aibjckdl} &=& g_1^{aibjckdl} - g_2^{aibjckdl}  \\
\nonumber g^{aibjckdl} &=& g_3^{(+)aibjckdl} + g_3^{(-)aibjckdl}.
\end{eqnarray} 
Let us focus in detail on the first case.  The structure constants are 
\begin{equation}\label{Neq5g}
g^{aibjckdl} = (\delta^{ac}\delta^{bd} - \delta^{ad}\delta^{bc})J^{ij}J^{kl} - \delta^{ab}\delta^{cd}(J^{ik}J^{jl} + J^{jk}J^{il}).
\end{equation}
They give rise to the following gauge transformation:
\begin{eqnarray}
\delta Z^{Adl} &=& -\frac{3i}{2}\bar{\xi}^{DF}_{[2}\xi_{1]BF}\omega_{DC} Z_{bk}^{B} Z^{Cl}_{b}Z^{Adk}  \\
\nonumber &&- \frac{3i}{2} \bar{\xi}^{DF}_{[2}\xi_{1]BF}\omega_{DC} Z_{b}^{Bk} Z^{Cd}_k Z^{Al}_b  .
\end{eqnarray}
The two terms are Sp(2$N$) and SO($M$) transformations, respectively, with matter fields in the fundamental representations of each \cite{Aharony:2008gk,Bergshoeff:2008bh,deMedeiros:2009eq}.\\

For the second case, the structure constants are simply
\begin{equation}\label{Neq5g2}
g^{aibjckdl} = J^{ik}J^{jl} \delta^{ac}\delta^{bd}  +  J^{il}J^{jk} \delta^{ad}\delta^{bc}.
\end{equation}
The indices are in standard direct product form, so the theory has gauge group Sp(2$MN$), with matter fields in the 2$MN$ dimensional fundamental representation.\\

For the special case of SO(4) $\times$ Sp(2) $\simeq$ SO(4) $\times$ SU(2), it is possible to add another term to the structure constants \cite{Bergshoeff:2008bh,deMedeiros:2009eq}:
\begin{equation}\label{SO(4)}
g^{aibjckdl} = g_1^{aibjckdl} - g_2^{aibjckdl}  + \alpha \varepsilon^{abcd} J^{ij} J^{kl},
\end{equation}
where $\varepsilon^{abcd}$ is the totally antisymmetric SO(4)-invariant tensor.  The resulting $g^{aibjckdl}$ satisfy (\ref{symm2}) and the fundamental identity, for any choice of the free parameter $\alpha$.  The gauge group closes into SO(4) $\times$ SU(2) for $\alpha \ne \infty$.  In the next section, we will see that this example, in the limit $\alpha \rightarrow \infty$, has gauge group SO(4).  In this limit, it lifts to ${\cal N} = 6$ and $8$. \\

There are also two ``exceptional" theories with ${\cal N} = 5$.  The first arises from the tensor
\begin{equation}
\label{G2structure}
g^{aibjckdl} = g_1^{aibjckdl} - g_2^{aibjckdl}  + \beta C^{abcd} J^{ij} J^{kl},
\end{equation}
where $a,b,... = 1,...\ 7$ and $i,j,... = 1,2$ are SO(7) and SU(2) indices, respectively.  Here $C^{abcd}$ is the totally antisymmetric tensor that is dual to the octonionic structure constants $C_{efg}$,
\begin{equation}
C^{abcd} = \frac{1}{3!}\varepsilon^{abcdefg}C_{efg}.
\end{equation}    
[For a concise introduction to G$_2$, SO(7) and the octonians, as well as a host of useful identities, see Section 2 and Appendix A of \cite{Bilal:2001an}.]   The tensor (\ref{G2structure}) satisfies (\ref{symm2}) and the fundamental identity for $\beta = 0$ or $\beta =  \frac{1}{2}$.  When $\beta=0$, the $g^{aibjckdl}$ are just the Sp(2) $\times$ SO(7) structure constants discussed above. \\

When $\beta = \frac{1}{2}$, the gauge group is G$_2$ $\times$ SU(2).  In this case, the structure constants take the form
\begin{equation}\label{G2g}
g^{aibjckdl} = (\delta^{ac}\delta^{bd} - \delta^{ad}\delta^{bc} + \frac{1}{2}C^{abcd})J^{ij}J^{kl} - \delta^{ab}\delta^{cd}(J^{ik}J^{jl}+J^{jk}J^{il }),
\end{equation}
where $i,j,... = 1,2$.  The gauge transformation is then
\begin{eqnarray}
\nonumber \delta Z^{Adl} &=& \tilde{\Lambda}^{aidl}Z^A_{ai},
\end{eqnarray}
with
\begin{eqnarray}
\tilde{\Lambda}^{aidl} &=& \frac{3i}{2} \bar{\xi}^{DF}_{[2}\xi_{1]BF} \omega_{DC} \delta^{ad} Z^{Bi}_b Z^{Cl}_b  \\
\nonumber && -\frac{3i}{4} \bar{\xi}^{DF}_{[2}\xi_{1]BF} \omega_{DC} (\delta^{ab}\delta^{cd} - \delta^{ac}\delta^{bd} + \frac{1}{2}C^{abcd})J^{jk}J^{il}Z^B_{bj}Z^C_{ck}.
\end{eqnarray}
The first term is clearly an SU(2) transformation.  The second is a G$_2 \subset$ SO(7) transformation, as can be seen by recognizing that the operator
\begin{equation}
{\cal P}^{abcd}_{14} = \frac{1}{3}\left(\delta^{ab}\delta^{cd} - \delta^{ac}\delta^{bd} + \frac{1}{2}C^{abcd}\right)
\end{equation}
is a projector from the adjoint $\bf 21$ of SO(7) to the adjoint $\bf 14$ of G$_2$, 
\begin{equation}
{\cal P}^{abcd}_{14}C_{bce} = 0.
\end{equation}
In this way we construct the ${\cal N} = 5,$ G$_2$ $\times$ SU(2) gauge theory from a 3-algebra, recovering the result found in \cite{Bergshoeff:2008bh,deMedeiros:2009eq}.   \\

The second exceptional theory has SO(7) $\times$ SU(2) gauge symmetry with matter transforming in the {\it spinor} $\bf 8$ of SO(7) \cite{Bergshoeff:2008bh,deMedeiros:2009eq}.  To find the structure constants, we start with the tensor
\begin{equation}
g^{aibjckdl} =  \delta^{ab}\delta^{cd}(J^{ik} J^{jl} +J^{jk} J^{il})+
\gamma \Gamma^{ab}_{mn}\Gamma^{cd}_{mn} J^{ij} J^{kl}.
\end{equation} 
where $a,b,... = 1,...\ 8$ and $i,j,... = 1,2$, and  $\Gamma^{ab}_{mn} = \frac{1}{2}(\Gamma_m\Gamma_n - \Gamma_n\Gamma_m)^{ab}$ is built from the $SO(7)$ gamma matrices.  The $g^{aibjckdl}$ have the correct symmetries and satisfy the fundamental identity for $\gamma = -\frac{1}{6}$, in which case the structure constants become
\begin{equation}
g^{aibjckdl} = \delta^{ab}\delta^{cd}(J^{ik}J^{jl} + J^{jk}J^{il})-
\frac{1}{6} \Gamma^{ab}_{mn}\Gamma^{cd}_{mn}J^{ij}J^{kl}.
\end{equation} 
The gauge transformations reduce to
\begin{eqnarray}
\delta Z^{Adl} &=& \tilde{\Lambda}^{aidl}Z^A_{ai},
\end{eqnarray}
where
\begin{eqnarray}
\tilde{\Lambda}^{aidl} &=& -\frac{3i}{2} \bar{\xi}^{DF}_{[2}\xi_{1]BF} \omega_{DC} \delta^{ad} Z^{Bi}_b Z^{Cl}_b  \\
\nonumber && +\frac{i}{8} \bar{\xi}^{DF}_{[2}\xi_{1]BF} \omega_{DC} 
 \Gamma^{ad}_{mn}\Gamma^{bc}_{mn}
J^{jk}J^{il}Z^B_{bj}Z^C_{ck}.
\end{eqnarray}
We see that the gauge group is SO(7) $\times$ SU(2), with the matter fields in the spinor representation of each.

\section{\sl Lifts: ${\cal N} = 5 \rightarrow {\cal N} = 6$}

In this section, we lift two theories with ${\cal N} = 5$ supersymmetry to ${\cal N} = 6$, along the lines of the lift from ${\cal N} = 6$ to ${\cal N} = 8$.  In particular, we lift the ${\cal N} = 5$ theories with Sp($2N$) $\times$ SO(2) and SO(4) $\times$ SU(2) gauge symmetry to ${\cal N} = 6$ theories with Sp($2N$) $\times$ U(1) and SO(4) gauge symmetry, respectively.  As we showed previously, the latter theory can then be lifted to ${\cal N} = 8$.  \\

To carry out the lifts, we first define unconstrained complex-conjugate scalars ${\cal Z}^A_a$ and $\bar{\cal Z}_A^a$, consistent with the constraint (\ref{constraint}):
\begin{eqnarray}\label{calZ}
{\cal Z}^A_a &=& Z^A_{a1} + iZ^A_{a2} \nonumber \\[2mm]
 \bar{{\cal Z}}_A^a &=& \bar Z_A^{a1} - i\bar Z_A^{a2}.
\end{eqnarray}
Supersymmetry then requires that the superpartner $\Xi_{Aa}$ be defined as follows:
\begin{eqnarray}\label{calPsi}
\Xi_{Aa} &=& \Psi_{Aa1} + i\Psi_{Aa2} \nonumber \\[2mm]
\Xi^{*Aa} &=& \Psi^{Aa1} - i\Psi^{Aa2}.
\end{eqnarray}
The indices 1 and 2 refer to either SU(2) or SO(2), while $a$ refers to SO(4) or Sp(2$N$), respectively.  The constraint (\ref{constraint}) allows us to write the complex-conjugate expressions in terms of the original fields.  Note that this procedure only works when one of the ${\cal N}=5$ gauge groups is SU(2) or SO(2). \\

We first consider the theory with Sp($2N$) $\times$ SO(2) gauge symmetry, where $a,b,...=1,...\ 2N$ are Sp($2N$) indices, and $i,j,... = 1,2$ are SO(2).  The conjugate scalar $\bar{{\cal Z}}_{A}^{a}$ takes the form
\begin{equation}
\bar{{\cal Z}}_{A}^{a} = -\omega_{AB} J^{ab}(Z^B_{b1}-i Z^B_{b2}), 
\end{equation}
and likewise for the conjugate spinor $\Xi^{*Aa}$.
With these definitions, it is straightforward to check that the ${\cal N} = 5$ transformations, with 
\begin{equation}
g^{aibjckdl} = -\frac{2}{3}\left((\delta^{ik}\delta^{jl} - \delta^{il}\delta^{jk})J^{ab}J^{cd} - \delta^{ij}\delta^{kl}(J^{ac}J^{bd} + J^{bc}J^{ad})\right),
\end{equation}
coincide with the ${\cal N} = 6$ transformations, with
\begin{equation}
f^{ab}{}_{cd} = J^{ab} J_{cd} + (\delta^a _c \delta^b _d - \delta^a _d \delta^b _c ),
\end{equation} 
for five of the six supersymmetries.  \\

To find the sixth, we plug $\varepsilon_{AB} \rightarrow -i \omega_{AB} \eta$ into the transformations (\ref{Neq6trans}) and collect terms.  After some calculation, we find:
\begin{eqnarray}
\nonumber \delta Z^A_{dl} &=& - \omega^{AD} \bar\eta \Psi_{Ddl} \\[1mm]
\nonumber
\delta\Psi_{Ddl} &=& -i\gamma^{\mu}\omega_{AD} \eta D_{\mu}Z^A_{dl}  \nonumber \\[2mm]
&&+\ i f^{ab}{}_{cd}(\omega_{AB}\omega_{CD} -\omega_{AC}\omega_{BD})\nonumber
\\[2mm]
&&\times (\epsilon_{ik}\epsilon_{jl} + \epsilon_{jk}\epsilon_{il} + i \delta_{ij}\epsilon_{kl})
Z^A_{ai}Z^B_{bj}Z^{Cc}_{k}\, \eta \nonumber \\[1mm]
\delta \tilde{A}_{\mu}{}^{aidl} &=& i f^{abcd} (\bar{\eta}\gamma _{\mu} \Psi_{Bbj}Z^B_{ck}
-\bar{\eta}\gamma _{\mu} \Psi_{Bck}Z^B_{bj}) (\delta^{jk}\epsilon^{il}+\epsilon^{jk}\delta^{il}),
\end{eqnarray} 
where $\epsilon^{ij}$ is the antisymmetric, invariant tensor of SO(2).  This is the extra supersymmetry transformation that lifts the ${\cal N} = 5$ theory with Sp($2N$) $\times$ SO(2) gauge symmetry to the ${\cal N} = 6$ theory with Sp($2N$) $\times$ U(1).  \\

Finally, we consider the ${\cal N} = 5$ theory with SO(4) $\times$ SU(2) gauge symmetry, with $g^{aibjckdl}$ given in (\ref{SO(4)}), in the limit $\alpha \rightarrow \infty$.  In this limit, the structure constants reduce to
\begin{equation}
g^{aibjckdl}  \rightarrow \alpha \varepsilon^{abcd} \epsilon^{ij} \epsilon^{kl},
\end{equation} 
where $a,b,...=1,...\ 4$ are SO(4) indices, and $i,j,... = 1,2$ are SU(2), and $\epsilon^{ij}$ is the antisymmetric, invariant tensor of SU(2).  We first compute the gauge transformation.  Using (\ref{Neq5gauge}), we find
\begin{equation}
\delta Z^{D}_{dl}\ \propto\ \bar{\xi}^{EF}_{[2}\xi_{1]BF}\omega_{EC} \varepsilon^{abcd} \epsilon^{jk} Z^B_{bj}Z^C_{ck}Z^D_{al}.
\end{equation}
This is a pure SO(4) gauge transformation; it suggests that the SO(4) $\times$ SU(2) invariant ${\cal N}=5$ theory, in the $\alpha \rightarrow \infty$ limit, can be lifted to the SO(4) theory with ${\cal N}=6$ and 8.\\

We now construct the lift.  We start by defining the complex-conjugate scalars ${\cal Z}^{A}_{a}$ and $\bar{{\cal Z}}_{A}^{a}$.  For the case at hand, we find
\begin{equation}
\bar{{\cal Z}}_{Aa} = -i\omega_{AB}(Z^B_{a1}-i Z^B_{a2}),
\end{equation}
and likewise for the spinor $\Xi^{*Aa}$.
As above, it possible to show that the ${\cal N} = 5$ transformations with 
\begin{equation}
g^{aibjckdl} = -\frac{2}{3} \varepsilon^{abcd} \epsilon^{ij} \epsilon^{kl},
\end{equation}
and the ${\cal N} = 6$ transformations with
\begin{equation}
f^{abcd} = \epsilon^{abcd},
\end{equation} 
coincide for five of the six supersymmetries.  \\

The sixth supersymmetry is derived in the same way as before.  Plugging $\varepsilon_{AB} \rightarrow -i \omega_{AB} \eta$ into (\ref{Neq6trans}) and collecting terms, we find:
\begin{eqnarray}
\nonumber \delta Z^A_{dl} &=& - \omega^{AD} \bar\eta \Psi_{Ddl} \\[1mm]
\nonumber
\delta\Psi_{Ddl} &=& -i\gamma^{\mu}\omega_{AD} \eta D_{\mu}Z^A_{dl}  \nonumber \\[2mm]
&&+\ 2 \varepsilon^{abcd} \, \omega_{AB}\omega_{CD}\, \delta_{ik}\delta_{jl} \,
Z^A_{ai}Z^B_{bj}Z^C_{ck}\, \eta \nonumber \\[1mm]
\delta \tilde{A}_{\mu}{}^{aidl} &=& -2i  \varepsilon^{abcd} \epsilon^{il} \bar{\eta}\gamma _{\mu} \Psi_{Bbj} Z^{B}_{cj}.
\end{eqnarray}
Note that the interaction term explicitly breaks the SU(2) symmetry.  The transformation is just what we need to lift the ${\cal N} = 5$ theory with $SU(2)\times SO(4)$ gauge symmetry to the ${\cal N} = 6$ theory with $SO(4)$ gauge symmetry.  In Section \ref{6to8}, we proved that this theory can again be lifted to ${\cal N} = 8$.  \\

It is worth emphasizing that these lifts arise from ${\cal N} = 5$ theories that are not simply ${\cal N} = 6$ theories with a reality constraint.  Instead they arise from purely ${\cal N} = 5$ theories, using very special properties of the gauge groups in question.

\section{\label{sec:level6}\sl Conclusions}

In this paper, we constructed the most general three-dimensional ${\cal N} = 5$ superconformal Chern-Simons gauge theory from first principles.  We identified the 3-algebra, found the fundamental identity, and constructed various representations of it.  We used 3-algebras to demonstrate how certain theories can be lifted to ${\cal N} = 6$ or 8 for an appropriate choice of gauge group.  \\

Our results confirm that 3-algebras provide a powerful approach to\break superconformal Chern-Simons theories in three dimensions \cite{deMedeiros:2009eq}.  They unify and simplify the construction of theories with ${\cal N} \ge 5$.  The number of supersymmetries is determined by the structure of the underlying 3-algebra.  Antisymmetric structure constants, with $f^{abcd} = -f^{bacd} = f^{cdab}$, give rise to ${\cal N} = 6$ theories, corresponding to U($M|N$) and OSp(2$|N$) in the Kac classification \cite{Kac:1977em}.  Symmetric structure constants, with $g^{abcd} = g^{bacd} = g^{cdab}$, give ${\cal N} = 5$ theories, corresponding to OSp($M|N$), D($2|1;\alpha$) and the exotic pair F(4) and G(3). \\

Perhaps our most surprising result is that theories with different gauge groups can be continuously connected through their 3-algebras.  How does this occur in an M2 brane construction?  We have seen that the ${\cal N} = 5$ supersymmetric SO(4) $\times$ SU(2) theory can be continuously deformed to the ${\cal N} = 6$ SO(4) theory, changing both gauge group and the number of supersymmetries along the way.  It is surely of interest to find the M theory realization of this phenomenon.\\

\section*{\sl Acknowledgments}

We would like to thank Neil Lambert, Steve Naculich, Raman Sundrum, and especially Arthur Lipstein for helpful discussions during the course of this work.  We would also like to thank Andreas Gustavsson, Jos\'e Figueroa-O'Farrill, Sung-Soo Kim, Paul de Medeiros, and Jakob Palmkvist for correspondence and questions that exposed a serious error in a previous version of this work.\\

This work was supported in part by the U.S. National Science Foundation, grant NSF-PHY-0401513.

\section*{\sl Appendix}

The theories we consider are constructed in three dimensions, with $\gamma^{\mu} = \{i\sigma^2 , \sigma^1 , \sigma^3\}$, with Minkowski metric $\eta^{\mu\nu} = (-,+,+)$.  Therefore, $\{\gamma^{\mu},\gamma^{\nu} \}= +2\eta^{\mu\nu}$.
In three dimensions, the Fierz transformation is
\begin{equation}
(\bar\lambda\chi)\psi = -\frac{1}{2}(\bar\lambda\psi)\chi
-\frac{1}{2} (\bar\lambda\gamma_\nu\psi)\gamma^\nu\chi .
\end{equation}
We use the symmetrization conventions $F^{A[B}G^{C]D} = F^{AB}G^{CD} - F^{AC}G^{BD}$ and  $F^{A(B}G^{C)D} = F^{AB}G^{CD} + F^{AC}G^{BD}$, for any parameters $F,G$, and indices $A,B,C,D$.  We adopt the convention $X^A = \omega^{AB} X_B$, $X_A = - \omega_{AB} X^B$ for any symplectic structure.  \\

Throughout the paper, we denote spinors that are R-symmetry singlets by $\eta$.  Those in the ${\bf 6}$ of SU(4) are denoted by $\varepsilon^{AB}$; those in the ${\bf 5}$ of Sp(4) are denoted by $\xi^{AB}$.  We note the following useful identities, which hold for both $\varepsilon^{AB}$ and $\xi^{AB}$, although they are presented for the latter, with the appropriate definition of $\varepsilon^{ABCD}$:
\begin{eqnarray}
\frac{1}{2}\bar\xi^{CD}_1\gamma_\nu\xi_{2CD}\,\delta^A_B&=&\bar\xi^{AC}_{[1}\gamma_\nu\xi_{2]BC}
 \end{eqnarray}
\begin{eqnarray} 
2\bar\xi^{AC}_{[1}\xi_{2]BD} &=&\bar\xi^{CE}_{[1}\xi_{2]DE}\delta^A_B-\bar\xi^{AE}_{[1}\xi_{2]DE}\delta^C_B \nonumber \\
&&+\ \bar\xi^{AE}_{[1}\xi_{2]BE}\delta^C_D - \bar\xi^{CE}_{[1}\xi_{2]BE}\delta^A_D
\end{eqnarray}
\begin{eqnarray}
\nonumber \frac{1}{2}\varepsilon_{ABCD}
\bar\xi^{EF}_1\gamma_\mu\xi_{2EF}&=&\bar\xi_{AB[1}\gamma_\mu\xi_{2]CD}+\bar\xi_{AD[1}\gamma_\mu\xi_{2]BC} \\
&&-\ \bar\xi_{BD[1}\gamma_\mu\xi_{2]AC}
\end{eqnarray}
\begin{eqnarray}
\varepsilon^{ABCD} &=& \omega^{AC}\omega^{BD}  - \omega^{AD}\omega^{BC}-\ \omega^{AB}\omega^{CD}.
\end{eqnarray}\\[-5mm]

In our calculations concerning $G_2$ and the spinor representation of $SO(7)$, we made considerable use of the representations and identities listed in \cite{Bilal:2001an}. The $SO(7)$ gamma matrices are
\begin{equation}
\Gamma^{mab} = i(C^{mab} + \delta^{ma}\delta^{b8} - \delta^{mb}\delta^{a8}).
\end{equation}
They lead to the $SO(7)$ generators
\begin{equation}
\Gamma^{mnab} = C^{mnab} + C^{mna}\delta^{b8} - C^{mnb}\delta^{a8} + \delta^{ma}\delta^{nb} - \delta^{mb}\delta^{na},
\end{equation}
which require the following $SO(7)$ identities:
\begin{eqnarray}
C^{abe}C^{cde} &=& -C^{abcd} + \delta^{ac}\delta^{bd} - \delta^{ad}\delta^{bc} \\
C^{acd}C^{bcd} &=& 6\delta^{ab} \\
C^{abpq}C^{pqc} &=& -4C^{abc}.
\end{eqnarray}
The $C^{abc}$ are the structure constants for the octonian algebra, and
\begin{equation}
C^{abcd} = \frac{1}{3!}\varepsilon^{abcdefg}C_{efg}.
\end{equation}


\newpage

\begin{thebibliography}{10}

\bibitem{Bagger:2006sk}
  J.~Bagger and N.~Lambert,
  ``Modeling multiple M2's,''
  Phys.\ Rev.\  D {\bf 75}, 045020 (2007)
  [arXiv:hep-th/0611108].

\bibitem{Bagger:2007jr}
  J.~Bagger and N.~Lambert,
  ``Gauge Symmetry and Supersymmetry of Multiple M2-Branes,''
  Phys.\ Rev.\  D {\bf 77}, 065008 (2008)
  [arXiv:0711.0955 [hep-th]].

\bibitem{Bagger:2007vi}
  J.~Bagger and N.~Lambert,
  ``Comments On Multiple M2-branes,''
  JHEP {\bf 0802}, 105 (2008)
  [arXiv:0712.3738 [hep-th]].

\bibitem{Gustavsson:2007vu}
  A.~Gustavsson,
  ``Algebraic structures on parallel M2-branes,''
  Nucl.\ Phys.\  B {\bf 811}, 66 (2009)
  [arXiv:0709.1260 [hep-th]].

\bibitem{Basu:2004ed}
  A.~Basu and J.~A.~Harvey,
  ``The M2-M5 brane system and a generalized Nahm's equation,''
  Nucl.\ Phys.\  B {\bf 713}, 136 (2005)
  [arXiv:hep-th/0412310].

\bibitem{Schwarz:2004yj}
  J.~H.~Schwarz,
  ``Superconformal Chern-Simons theories,''
  JHEP {\bf 0411}, 078 (2004)
  [arXiv:hep-th/0411077].

\bibitem{Aharony:2008ug}
  O.~Aharony, O.~Bergman, D.~L.~Jafferis and J.~Maldacena,
   ``N=6 superconformal Chern-Simons-matter theories, M2-branes and their
  gravity duals,''
  JHEP {\bf 0810}, 091 (2008)
  [arXiv:0806.1218 [hep-th]].

\bibitem{Aharony:2008gk}
  O.~Aharony, O.~Bergman and D.~L.~Jafferis,
  ``Fractional M2-branes,''
  JHEP {\bf 0811}, 043 (2008)
  [arXiv:0807.4924 [hep-th]].
  
\bibitem{Hosomichi:2008jb}
  K.~Hosomichi, K.~M.~Lee, S.~Lee, S.~Lee and J.~Park,
  ``N=5,6 Superconformal Chern-Simons Theories and M2-branes on Orbifolds,''
  JHEP {\bf 0809}, 002 (2008)
  [arXiv:0806.4977 [hep-th]].

\bibitem{Schnabl:2008wj}
  M.~Schnabl and Y.~Tachikawa,
  ``Classification of N=6 superconformal theories of ABJM type,''
  arXiv:0807.1102 [hep-th].

\bibitem{Bagger:2008se}
  J.~Bagger and N.~Lambert,
  ``Three-Algebras and N=6 Chern-Simons Gauge Theories,''
  Phys.\ Rev.\  D {\bf 79}, 025002 (2009)
  [arXiv:0807.0163 [hep-th]].

\bibitem{Lambert:2010ji}
  N.~Lambert and C.~Papageorgakis,
 ``Relating U(N)xU(N) to SU(N)xSU(N) Chern-Simons Membrane theories,''
  arXiv:1001.4779 [hep-th].
  
\bibitem{deMedeiros:2008zh}
  P.~de Medeiros, J.~Figueroa-O'Farrill, E.~Mendez-Escobar {\it et al.},
``On the Lie-algebraic origin of metric 3-algebras,''
  Commun.\ Math.\ Phys.\  {\bf 290}, 871-902 (2009)
  [arXiv:0809.1086 [hep-th]].
   


\bibitem{Bergshoeff:2008bh}
  E.~A.~Bergshoeff, O.~Hohm, D.~Roest, H.~Samtleben and E.~Sezgin,
  ``The Superconformal Gaugings in Three Dimensions,''
  JHEP {\bf 0809}, 101 (2008)
  [arXiv:0807.2841 [hep-th]].
  
\bibitem{Chen:2009cwa}
  F.~M.~Chen,
  ``Symplectic Three-Algebra Unifying N=5,6 Superconformal Chern-Simons-Matter
  Theories,''
  JHEP {\bf 1008}, 077 (2010)
  [arXiv:0908.2618 [hep-th]].
  

\bibitem{Bergshoeff:2008cz}
  E.~A.~Bergshoeff, M.~de Roo and O.~Hohm,
  ``Multiple M2-branes and the Embedding Tensor,''
  Class.\ Quant.\ Grav.\  {\bf 25}, 142001 (2008)
  [arXiv:0804.2201 [hep-th]].

\bibitem{Bergshoeff:2008ix}
  E.~A.~Bergshoeff, M.~de Roo, O.~Hohm and D.~Roest,
  ``Multiple Membranes from Gauged Supergravity,''
  JHEP {\bf 0808}, 091 (2008)
  [arXiv:0806.2584 [hep-th]].
  
\bibitem{deMedeiros:2009eq}
  P.~de Medeiros, J.~Figueroa-O'Farrill and E.~Mendez-Escobar,
  ``Superpotentials for superconformal Chern-Simons theories from
  J.\ Phys.\ A  {\bf 42}, 485204 (2009)
  [arXiv:0908.2125 [hep-th]].


\bibitem{Wess:1992cp}
  J.~Wess and J.~Bagger,
{\it Supersymmetry and Supergravity,}
(Princeton University Press, 1992).


\bibitem{Bilal:2001an}
  A.~Bilal, J.~P.~Derendinger and K.~Sfetsos,
 ``(Weak) G(2) holonomy from self-duality, flux and supersymmetry,''
  Nucl.\ Phys.\  B {\bf 628}, 112 (2002)
  [arXiv:hep-th/0111274].

\bibitem{Kac:1977em}
  V.~G.~Kac,
  ``Lie Superalgebras,''
  Adv.\ Math.\  {\bf 26}, 8 (1977).
  
\end{thebibliography}
\end{document}